\documentstyle[12pt]{article}
\newcommand{\doublespacing}{\let\CS=\@currsize\renewcommand{\baselinesstrech}
{2.0}\tiny\CS}

\begin{document}

\title{A Lie algebraic approach to complex quasi exactly solvable potentials
with real spectrum}
\author{P.Roy.\thanks{E-mail : pinaki@isical.ac.in} and
R.Roychoudhury\thanks{E-mail : raj@isical.ac.in}\\ Physics \& Applied
Mathematics Unit\\ Indian Statistical Institute \\ Calcutta  700035\\ India} 


\maketitle

\vspace*{1.5cm}

\centerline{\bf Abstract}

\vspace{0.3cm}

\thispagestyle{empty}

\setlength{\baselineskip}{18.5pt}
We use a Lie algebraic technique to construct complex quasi exactly solvable 
potentials with real spectrum. In particular we obtain exact solutions of a
complex sextic oscillator potential and also a complex potential belonging to
the Morse family.

\newpage

Since the paper of Bender and Boetcher \cite{ben} there has been a great deal of 
interest in the study of non hermitian PT-symmetric potentials with real spectrum 
\cite{ben1,can,zno,zno1,bag}. Recently a particularly interesting class of
PT-symmetric polynomial
potential has also been found \cite{bag1}. In the present paper we
shall use a method based on Lie algebra \cite{tur,shif,ush} to construct
quasi exactly solvable complex sextic oscillator potential with real spectrum.
We shall also construct another potential of the Morse family with the same
property.      

To begin with we consider the Schr\"odinger equation
\begin{equation}
H\psi = [-\frac{d^2}{dx^2} + V(x)]\psi = E\psi \label{sch}
\end{equation}
where the potential $V(x)$ is given by
\begin{equation}
V(x) = x^6 + \alpha x^4 + \beta x^2  
\end{equation}
We now make the following transformation
\begin{equation}
\begin{array}{lcl}
\psi(x) &=& exp[-\int W(x) dx] \phi (x)\\ \\
W(x) &=& x^3 + ax \footnotemark \\ \label{W}
\end{array}
\end{equation}
\footnotetext{In what follows we shall obtain even parity solutions.
To obtain odd parity solutions one should add a term $1/x$ to $W(x)$.}
and obtain from equation (\ref{sch}) 
\begin{equation}
-[\frac{d^2}{dx^2} - 2W(x)\frac{d}{dx} + V(x) - W^2(x) + W'(x) - E]\phi = 0 \label {modi}
\end{equation}

Next we perform the substitution
\begin{equation}
z = x^2
\end{equation}
and obtain from equation (\ref{modi})
\begin{equation}
-4z \frac{d^2\phi}{dz^2} + (4z^2 + 4az - 2) \frac{d\phi}{dz} + [(\alpha-2a)z^2 + (3+\beta-a^2)z + a - E]\phi = 0 \label{modi1}
\end{equation}

We now consider the opeartors $J^+ , J^0$ and $J^-$  given by
\begin{equation}
J^+ = z^2 \frac{d}{dz} - 2jz~~,~~J^0 = z \frac{d}{dz} - j~~,~~J^- = \frac{d}{dz} \label{generators}
\end{equation}
It is easy to verify that $J^+,J^0$ and $J^-$ satisfy the Su(2) algebra :
\begin{equation}
[J^+ , J^-] = -2J^0~~,~~[J^0 , J^\pm] = \pm J^\pm
\end{equation}
Thus $J^+,J^0$ and $J^-$ constitute a finite dimensional representation of
the sl(2) group with spin j in a space of polynomials with basis functions 
$z^{j+m} (-j \leq m \leq j)$ \cite{tur,shif}.

It is now necessary to write equation (\ref{modi1}) in terms of the generators
in (\ref{generators}). To do this we consider the following expression :
\begin{equation}
[A J^0J^- + BJ^+ + CJ^- + DJ^0 + \epsilon] \phi = 0 \label{modi2}
\end{equation}
Using (\ref{generators}) we obtain from (\ref{modi2})
\begin{equation}
Az\frac{d^2\phi}{dz^2} + (Bz^2 + Dz - Aj + c)\frac{d\phi}{dz} + (\epsilon - Dj - 2Bjz)\phi = 0 \label{modi3}
\end{equation}
Now identifying (\ref{modi1}) and (\ref{modi3}) we obtain
\begin{equation}
\begin{array}{lcl}
A &=& -4~~,~~B = 4~~,~~C = -(2+4j)~~,~~D = 4a\\ \\
\alpha &=& 2a~~,~~\beta = \frac{\alpha^2}{4} - 2Bj - 3~~,~~\epsilon = Dj + a - E\\ \label{constraint} 
\end{array}
\end{equation}
Next using the relations (\ref{constraint}) we find that
\begin{equation}
V(x) = x^6 + 2ax^4 + (a^2-8j-3)x^2 \label{potential}
\end{equation}
We note that the parameter a appearing in (\ref{potential}) is a free 
parameter and various choices are possible. But for the sake of convenience
we choose $a=i\mu$ such that $\mu^2 < 2$. We shall now determine eigenvalues
and eigenfunctions corresponding to various values of $j$, keeping in mind
that $\phi$ is of the form $\phi = (1+c_1z+c_2z^2+....+c_{2j}z^{2j})$.

{\bf Case 1}:
Let $j = 0$. In this case we have $(2j+1) = 1$ solution with $\phi(z)=1$. After
a straight foroward calculation we find
\begin{equation}
\begin{array}{lcl}
V(x) &=& x^6 + 2i\mu x^4 - (\mu^2+3)x^2 + i\mu \\ \\
E &=& 0~~,~~\psi(x)\sim exp(-x^4/4 - i\mu x^2/2)\\ \label{wf1}
\end{array}
\end{equation}

{\bf Case 2}:
Let $j=1/2$. In this case we have $(2j+1) = 2$ solutions and the reduced wave
function is of the form $\phi(z) = (1 + c_1 z)$. As before we find that
\begin{equation}
\begin{array}{lcl}
V(x) &=& x^6 + 2i\mu x^4 - (\mu^2+7)x^2 - 3i\mu \\ \\  
E &=& -2\sqrt{2-\mu^2}~~,~~\psi(x)\sim [1+(\sqrt{2-\mu^2}-i\mu)x^2] exp[-x^4/4 - i\mu x^2/2]\\ \\
E &=& 2\sqrt{2-\mu^2} ~~,~~\psi(x)\sim [1-(\sqrt{2-\mu^2}-i\mu)x^2] exp[-x^4/4 - i\mu x^2/2]\\ \label{wf2}
\end{array}
\end{equation}
Now considering higher values of j i.e, $j=1,3/2,.....$ we can obtain potentials
admitting more exact solutions.

We now turn to a potential of the Morse family. The Schr\"odinger equation
is given by
\begin{equation}
[-\frac{d^2}{dx^2} + V(x) - E]\psi = 0 \label{morse}
\end{equation}
where the potential $V(x)$ is
\begin{equation}
V(x) = \alpha^2 e^{2x} + \beta e^x + \gamma e^{-x} + \delta^2 e^{2x}
\end{equation}
As in the previous example we use the transformation
\begin{equation}
\psi(x) = exp[-\int W(x) dx] \phi(x)
\end{equation}
with $W(x)$ given by 
\begin{equation}
W(x) = d e^x - a e^{-x} + b \label{w2}
\end{equation}
and the corresponding change of variable as 
\begin{equation}
z = e^{-x}
\end{equation}
Then the Schr\"odinger equation (\ref{morse}) becomes
\begin{equation}
-z^2 \frac{d^2\phi}{dz^2} + (2az^2 - 2bz - z - 2d) \frac{d\phi}{dz} + [(\gamma+2ab+a)z + 2ad - b^2 - E]\phi = 0 \label{modi4}
\end{equation}
where in obtaining (\ref{modi4}) we have taken
\begin{equation}
\alpha = d~~,~~\beta = d(2b-1)~~,~~\delta = a \label{iden2}
\end{equation}
Now we have to express the equation (\ref{modi4}) in terms of the combination
of the generators $J^+,J^0$ and $J^-$. It can be esily shown that this expression
is given by
\begin{equation}
[A J^+ J^- + B J^+ + C J^- + D J^0 + \epsilon]\phi = 0
\end{equation}
and using the explicit forms of $J^\pm,J^0$ we get
\begin{equation}
Az^2\frac{d^2\phi}{dz^2} + (Bz^2 - 2Ajz + Dz + C)\frac{d\phi}{dz} + (\epsilon-2Bjz-Dj)\phi = 0 \label{modi5}
\end{equation}
Now comparing (\ref{modi4}) and (\ref{modi5}) we find
\begin{equation}
\begin{array}{lcl}
A &=& -1~~,~~B = 2a~~,~~C = 2d~~,~~D = -(2b+1+2j)\\ \\
\gamma &=& -[2ab+(4j+1)a]~~,~~\epsilon = \frac{D}{2}+2ad-b^2-E\\ \label{iden3}
\end{array}
\end{equation}
Then using (\ref{iden2}) and (\ref{iden3}) the potential is found to be
\begin{equation}
V(x) = d^2 e^{2x} - d(1-2b) e^x - a(2b+4j+1) e^{-x} + a^2 e^{-2x}
\end{equation}
It is now possible to consider various cases depending on the values of $j$.
However,for the sake of simplicity,we shall consider the cases $j=0,1/2$. At
this point we shall make a convenient choice of the parameter $b$ and we take
$b=\frac{i\mu-3}{2}$ such that $16ad > \mu^2$.

{\bf Case 1}:
Let us take $j=0$. In this case there is one known solution.
\begin{equation}
\begin{array}{lcl}
V(x) &=& d^2 e^{2x} - d(4-i) e^x - a(i\mu-2) e^{-x} + a^2 e^{2x} - \frac{3i\mu}{2}\\ \\
E &=& 2ac - \frac{9-\mu^2}{4}~~,~~\psi(x) \sim exp[-de^x - ae^{-x} - \frac{i\mu-3}{2}x] \label{wf3}
\end{array}
\end{equation}

{\bf Case 2}:
In this case we take $j=1/2$ so that two solutions are known.
\begin{equation}
\begin{array}{lcl}
V(x) &=& d^2 e^{2x} - d(4-i) e^x - ia\mu e^{-x} + a^2 e^{-2x} - i\mu \\ \\ 
E &=& 2ad - \frac{9-\mu^2}{4} - \sqrt{16ad-\mu^2}~~,~~\psi(x) \sim (e^{-x}+\frac{-i\mu+\sqrt{16ad-\mu^2}}{4a})exp[-de^x - ae^{-x} - \frac{i\mu-3}{2}x]\\ \\
E &=& 2ad - \frac{9-\mu^2}{4} + \sqrt{16ad-\mu^2}~~,~~\psi(x) \sim (e^{-x}+\frac{-i\mu-\sqrt{16ad-\mu^2}}{4a})exp[-de^x - ae^{-x} - \frac{i\mu-3}{2}x]\\ \label{wf4}
\end{array}
\end{equation}
From (\ref{wf1}),(\ref{wf2}),(\ref{wf3}) and (\ref{wf4}) we find that in all the cases the energies
are real and the corresponding wave functions are normalisable. Another 
interesting fact is that in the all the four cases the potentials are complex
without being PT-symmetric. A similar situation has recently been reported
\cite{khare}.

It may also be noted that the relation between the formalism adopted in this
paper and supersymmetric quantum mechanics \cite{cooper} has previously been
noted \cite{roy,shifman}. It is not difficult to observe that the function
$W(x)$ (see for example eq(\ref{W})) can be identified with the superpotential
of supersymmetric quantum mechanics \cite{cooper} and the potentials considered
here are of the form $W^2(x)-W'(x)+ constant$. Since $W^2(x)+W'(x)$ is isospectral
with $W^2(x)-W'(x)$ it may be expected that by calculating $W^2(x)+W'(x)$ it
is possible to obtain other QES potentials which are complex but whose known
part of the spectrum is real. Finally we would like to mention that although
we have considered two types of potentials, it seems that all the potentials 
belonging to the sl(2) catalogue can be treated in a similar fashion.

\end{document}